\documentclass{jaa}

\usepackage{graphicx}

\begin{document}

\title{Whispers from the edge of physics}


\author{Nils Andersson}
\affilOne{Mathematical Sciences and STAG Research Centre, University of Southampton, UK.}


\twocolumn[{

\maketitle

\corres{na@maths.soton.ac.uk}


\begin{abstract}
Neutron stars involve extreme physics which is difficult (perhaps impossible) to explore in laboratory experiments. We have to turn to astrophysical observations, and try to extract information from the entire range of the electromagnetic spectrum. In addition, neutron stars may radiate gravitational waves through a range of scenarios. In this brief summary I outline some of the main ideas, focussing on what we do and do not know, and describe the challenges involved in trying to  catch these faint whispers from the very edge of physics.  \\

{\em Contribution to  Journal of Astrophysics and Astronomy special 
issue on 'Physics of Neutron Stars and related objects', celebrating
the 75th birth-year of G. Srinivasan.}
 \end{abstract}

\keywords{neutron stars---gravitational waves.}

}]


\doinum{10.1007/s12036-017-9463-8}
\artcitid{38:58}
\year{2017}

\section{Introduction}

If you pick up an iron ball and try to squeeze it you probably won't make much progress. The internal pressure of the metal easily withstands your push. However, imagine you had superhuman strength, then what would happen? As you squeeze the ball the internal pressure will increase until you reach the point where the ball is held up by the electron degeneracy, when the electron shells of the atoms "touch". If you keep pushing to overcome this pressure there is nothing stopping you until the atomic nuclei ``touch''. In the process, the material has changed to become more neutron rich and it is the neutron degeneracy pressure that provides the last defence against your godlike squeezing. Keep on pushing and you will  end up with your own mini black hole. 

Nature  reproduces the different stages of this thought experiment when stars die. As a normal star runs out of nuclear fuel it can no longer support itself against gravity so it begins to collapse. If the star is small enough, it compresses until it is supported by the electron degeneracy and a white dwarf is formed. However, Chandrasekhar taught us that white dwarfs have a maximum mass -- depending on composition, about $1.4M_\odot$. More massive remnants will continue to collapse until the neutron degeneracy comes into play. If this is strong enough, a neutron star is born. But neutron stars also have a maximum mass, so if the collapsing object is too heavy a black hole will form.

This is a simple enough story, but the details are complex. 
In fact, neutron stars represent many extremes of physics. With a mass of about one and a half times that of the Sun squeezed into a ball with radius of about 10 km (the size of a small city), the density reaches beyond what can be reproduced in our laboratories. In essence, the internal composition and state of matter are unknown. A complete description of a neutron star involves all four fundamental forces of nature. Gravity holds the star together. Electromagnetism makes it visible and the star's magnetic field also dictates the evolution of the spin rate. The strong interaction determines the internal composition, e.g. the number of protons per neutron, while weak interactions determine how rapidly the star cools (and also decide how viscous the internal fluid flow is). 

Neutron star modelling takes us to the edge of physics;  one must 
combine supranuclear physics with magnetohydrodynamics, a description of 
superfluids and superconductors, potentially exotic phases of matter like a 
deconfined quark-gluon plasma and, of course, general relativity. Moreover, one must aim to develop models that can explain a wide range of observed phenomena. 

Since the first radio pulsars were discovered 50 years ago, these enigmatic objects have primarily been probed by radio timing and X-ray timing and  spectra. However, there is more to neutron stars than you can see. 
They may also radiate  gravitational waves through a variety of scenarios, ranging from 
the supernova core collapse in which they are born to the merger of
binary systems. Mature neutron stars may radiate via asymmetries in their elastic crust or 
unstable waves (for example, associated with the inertial r-modes) in the fluid interior. Different scenarios
depend sensitively on specific aspects of neutron star physics (elasticity,
superfluidity, viscosity etc.), and the challenge is to i) model the mechanisms that generate the radiation in the first place, in order to facilitate template-based detection, and ii) hopefully decode observed signals
to ``constrain'' current theory. 

The celebrated LIGO detections of black-hole binary inspiral and merger  \cite{merger,merger2} demonstrate the discovery potential of gravitational-wave astronomy. As the sensitivity of the detectors improves, and a wider network of instruments comes online (including LIGO-India!), a broader range of sources should be detected. Neutron star signals are anticipated with particular excitement -- we are eagerly waiting for whispers from the edge of physics.

\section{Binary inspiral and merger}

Well before the first direct detection, we knew  Einstein had to be right. Precision radio timing of the orbital evolution of double neutron star systems, like the celebrated binary pulsar PSR1913+16, showed perfect agreement with the predicted energy loss due to gravitational-wave emission (to better than 1\%). Yet, this was not a test of the strong field aspects of general relativity. The two partners in all known binary neutron stars are so far apart that they can, for all intents and purposes, be treated as point particles (in a post-Newtonian analysis). The internal composition is immaterial. If we want to probe the involved matter issues we need to observe the late stages of inspiral. 

Double neutron star systems will spend their last 15 minutes or so in the sensitivity band of advanced ground-based interferometers (above 10 Hz). 
The detection of, and extraction of parameters from, such systems is of great importance for both astrophysics and nuclear physics. From the astrophysics point-of-view, observed event rates should lead to insights into the formation channel(s) for these systems and the identification of an electromagnetic counterpart to the merger should confirm the paradigm for short gamma-ray bursts. Meanwhile, the nuclear physics aspects relate to the equation of state for matter at supranuclear densities. 

\begin{figure}[h]
 \centerline{\includegraphics[width=9cm]{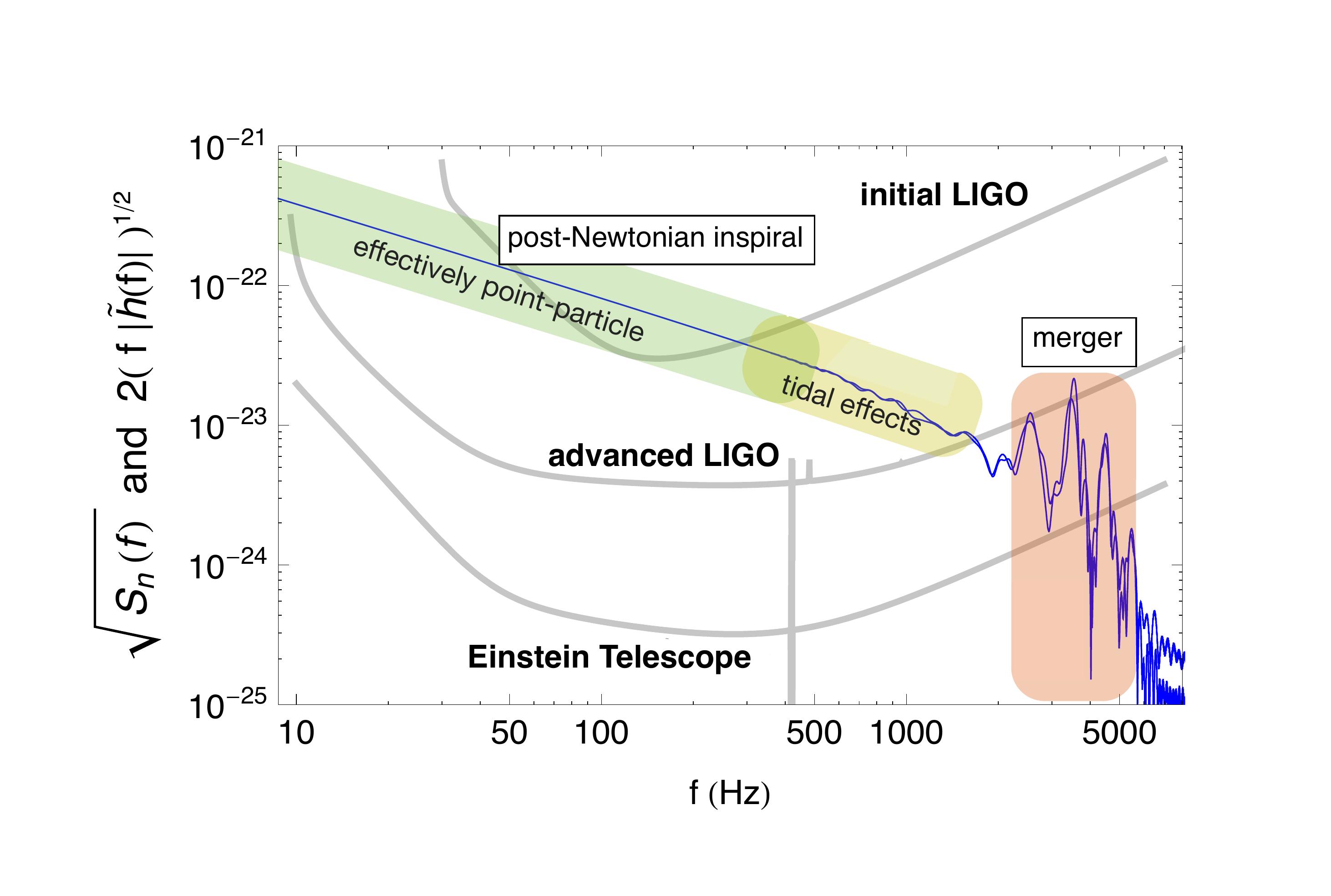}}
  \caption{A schematic illustration of the gravitational-wave signal emitted during the late stages of binary neutron star inspiral. The effective signal strain is compared to the sensitivity of different generations of detectors. Above 100~Hz or so the tidal compressibility is expected to leave a secular imprint on the signal. The eventual merger involves violent dynamics, which also encodes the matter equation of state. The merger signal is expected at a few kHz, making it difficult to observe with the current generation of detectors, but it should be within reach of third generation detectors like the Einstein Telescope. Adapted (with permission) from an original figure by J. Read (based on data from \cite{read}).}
\label{tides}
\end{figure}

Neutron star binaries allow us to probe the equation of state in unique ways, schematically illustrated in Figure~\ref{tides}. First of all, finite size effects come into play at some point during the system's evolution. An important question concerns to what extent the tidal interaction leaves an observable imprint on the gravitational-wave signal \cite{fh2008,th2010}. This problem has two  aspects. The tidal deformability of each star is encoded in the so-called Love numbers (which depend on the stellar parameters and represent the static contribution to the tide). This effect is typically expressed as
\begin{equation}
\lambda = {2 \over 3} k_2 R^5 \sim {\mathrm{quadrupole\  deformation} \over \mathrm{tidal\  field}}
\end{equation}
where $R$ is the star's radius and $k_2$ encodes the  compressibility of the stellar fluid. It is  difficult to alter the gravitational-wave phasing in an inspiralling binary  (as an example, an energy change of something like $10^{46}$~erg at 100 Hz  only leads to a shift of $10^{-3}$ radians), but the tidal deformation  may nevertheless lead to a distinguishable secular effect. Observing this effect will be challenging as we may need several tens of detections before we begin to distinguish between equations of state \cite{ag2015}. However, the strategy nevertheless promises to constrain the neutron star radius to better than 500~m. This could lead to stronger constraints on the equation of state than current and upcoming nuclear physics experiments. 

The star also responds dynamically to the tidal interaction.  As the binary sweeps through the detector's sensitive band a number of resonances with the star's oscillation modes may become relevant \cite{ks1995,hl1999}. In particular, it has recently been demonstrated that  \cite{steinhoff}-- even though it does not actually exhibit a resonance before the stars merge -- the tidal driving of the star's fundamental f-mode is likely to be significant (representing the dynamic tide). Quantifying these two effects may allow us to extract the stellar parameters  (both  mass and  radius for each of the two binary companions) and hence constrain the cold equation of state. 

These arguments presume that the tidal effects are weak. Provided this is the case, the main challenge is to extract precise information from observed signals. However, the tidal problem may be more complicated. It has recently been suggested \cite{wein2013,ess2016} that the (non-resonant) coupling between the tide and the star's p and g-modes may trigger an instability (when the system evolves beyond 50~Hz or so) that grows to the point where it has severe impact on the gravitational-wave signal -- potentially preventing  detection using current search templates. If this argument is correct it could have very serious implications. Unfortunately, it is far from easy to establish to what extent this is a real concern. The problem is difficult because the p-g instability involves very short wavelength oscillation modes, which depend sensitively on the internal physics. We may not have appropriate computational technology to resolve the issue. We clearly can not rely on numerical simulations, as the required resolution is way beyond what is feasible.  This is troublesome. If the p-g instability does play a role then our current numerical waveforms for binary neutron star signals would not represent reality and our search  templates would not be reliable. This would be very bad news, indeed.

In contrast, the eventual neutron star merger involves violent dynamics. As the two stars crash together, the fluid sloshes about violently. This leads to shocks which heat up the matter to a level beyond that of the supernova furnace in which the stars were first born. This involves composition changes and, as in the supernova core collapse problem, neutrinos  play an important role. Tracking the dynamics requires nonlinear simulations, with a live spacetime -- an extremely challenging problem. Nevertheless, this is an area that has seen impressive progress in the last few years (see \cite{baiotti} for a recent review).  Simulations are becoming more robust and the physics implementation more realistic. State-of-the-art simulations regularly include equations of state based on detailed nuclear physics calculations, electromagnetism (typically in the form of magnetohydrodynamics) and sometimes a  prescription for neutrino transport. However, including all these features involves an astonishing computational cost (a typical run would take months on the largest supercomputer you can get your hand on) and the limited achievable resolution means that it is difficult to distinguish local matter features (like the crust-core transition).

As the simulations become more sophisticated, we are learning that the merger event has identifiable features which depend (more or less) directly on the (hot) matter equation of state. In particular, peaks in the gravitational-wave spectrum can be identified with specific dynamical features. A particularly robust feature is associated with the star's f-mode \cite{ab,lr}. As the scaling of the f-mode frequency with the stellar parameters is fairly well understood one may be able to use detections to constrain the supranuclear physics. Unfortunately, the signal is expected at several kHz, see figure ~\ref{tides}, meaning that it may not be easily detected even with advanced LIGO/Virgo \cite{clark}. We may have to wait for third generation detectors, like the Einstein Telescope.

\section{Mountains}

Individual neutron stars may also be interesting gravitational-wave sources. Any rotating deformed body will radiate gravitationally, and in the case of neutron stars the required deformation can be
due to strain built up in the crust, the internal magnetic field or arise as a result of
accretion. 

For a triaxial star rotating steadily we have the raw gravitational-wave strain
\begin{equation}
h \approx 3\times10^{-28} \left( {\epsilon\over 10^{-6}}\right) \left({f_\mathrm{spin} \over 10~\mathrm{Hz}}\right)^2 \left({1~\mathrm{kpc} \over d}\right)
\end{equation}
where $\epsilon$ represents the (dimensionless) asymmetry in the moment of inertia tensor, $f_\mathrm{spin}$ is the rotation frequency and $d$ is the source distance. This is a text-book calculation. Unfortunately, it is difficult to make the real problem ``calculable'' because it involves poorly understood evolutionary aspects. Our current understanding  is mainly based
on attempts to work out the largest deformation the star can sustain,
e.g. before the crust breaks.  The best estimates come from molecular dynamics simulations, which suggest that the
crust is super-strong \cite{2009PhRvL.102s1102H}, and that it could, in principle, sustain
asymmetries as large as one part in $10^5$
\cite{2006MNRAS.373.1423H}.  However, this does not in any way suggest that neutron stars actually will have
deformations of this magnitude. Why would nature choose to deform stars to the  limit? 

As we struggle to make progress with the modelling, we may seek guidance from observations. 
The signal from a spinning neutron star is unavoidably weak, but the effective amplitude (after matched filtering) improves (roughly) as the square-root of the observation time. Given the expected maximum amplitude, we can easily work out that we need observations lasting at least one year. And we know the location and spin rate of many radio pulsars, so we have some idea of what we are looking for.  

So far, targeted pulsar searches may not have led to detections but the results are nevertheless interesting. An
observational milestone was reached when LIGO  used data from the first 9 months of the S5 science run  to beat the Crab pulsar spin-down limit \cite{ligo_crab}.  It may have been obvious from the beginning that there was no
real possibility that 100\% of the observed Crab pulsar spin-down was gravitational-wave
powered, as this would conflict with the measured braking index.  But the fact that the gravitational contribution to spin-down  is less than 0.2\% (as evidenced by the recent data \cite{o1} )  was not at all obvious.  We are also learning that millisecond pulsars have a high degree of perfection. The current record holder is PSR J0636+5129, which would radiate at $f_\mathrm{gw}\approx 697$~Hz (twice the spin frequency) and for which the current LIGO upper limit  is $\epsilon\approx 1.3\times10^{-8} $. This represents an astonishing level of symmetry.

\begin{figure}[h]
 \centerline{\includegraphics[width=8.5cm]{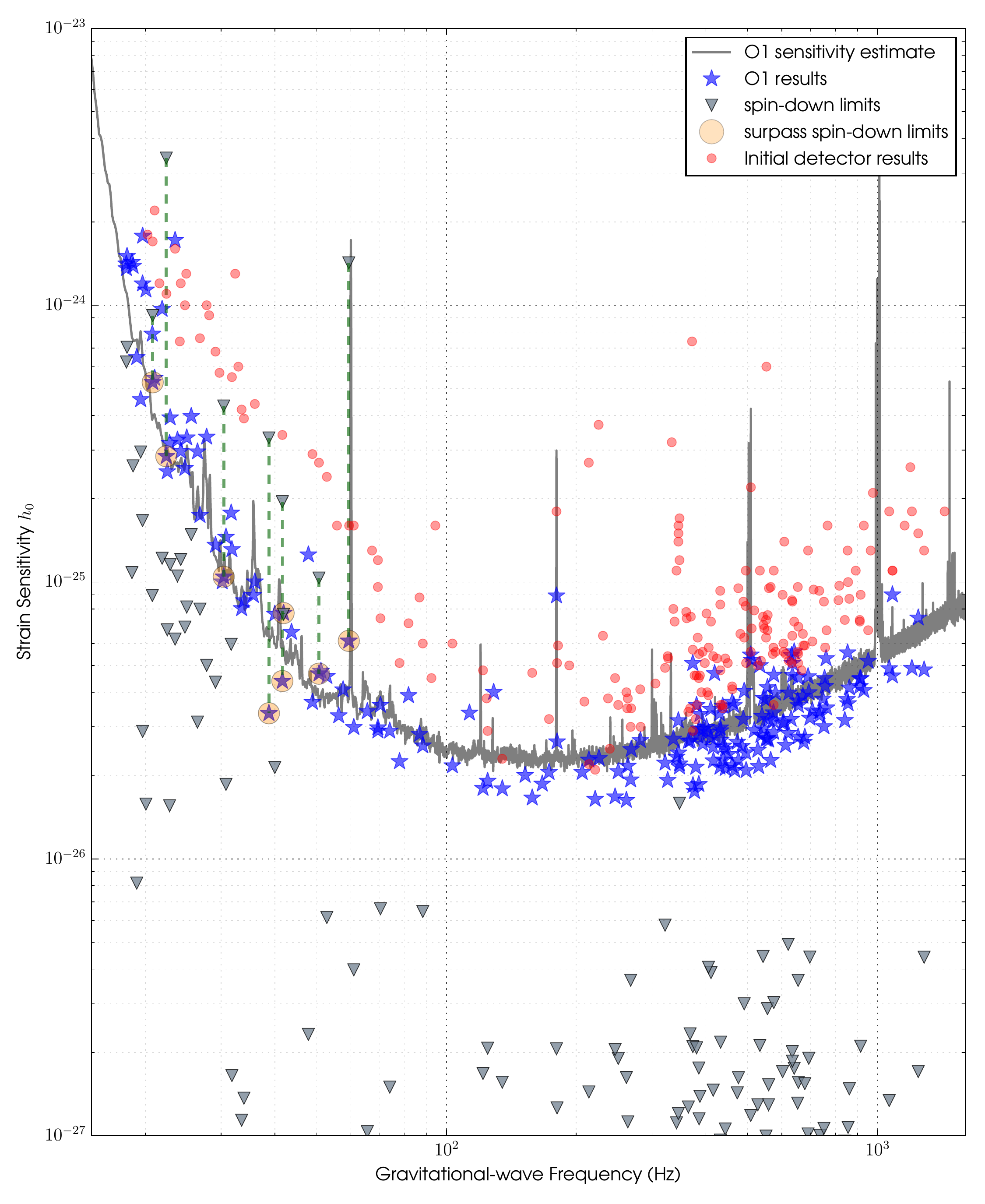}}
  \caption{Summary of targeted searches for gravitational waves from 200 pulsars from the first observing run of the Advanced LIGO detectors (O1).  The upside-down triangles give the spin-down limits for all (non-Globular Cluster) pulsars, based on values taken from the ATNF pulsar catalog and assuming a canonical moment of inertia. Stars show the observational upper limits, with  shaded circles indicating pulsars for which the spin-down limits (linked via the dashed vertical lines) 
were surpassed with the observations. The gray curve gives an estimate of the expected strain sensitivity for 
O1, combining representative amplitude spectral density for the two LIGO detectors. Reproduced from Abbott et al \cite{aasi}.}
\label{pulsars}
\end{figure}

This discussion highlights that the key question is not what the largest allowed deformation may be, but what the smallest one is. This is also tricky, but in this case we at least have a starting point.
The magnetic field will   deform the star, and as pulsars are magnetised this sets a lower limit. Unfortunately, this deformation is extremely small for typical
pulsar field strengths \cite{2008MNRAS.385..531H,2008MNRAS.385.2080C,2009MNRAS.395.2162L}:
\begin{equation}
\epsilon \approx 10^{-12} \left(\frac{B}{10^{12} \rm \, G}\right)^2 \ .
\end{equation}
Moreover, it is the \emph{internal}, rather than the external magnetic
field strength that counts.  This means that we have little guidance from the inferred external dipole field. We need the internal configuration and this is another tricky issue. For example, the above estimate assumes a normal
fluid core while real neutron stars are expected to harbour a proton superconductor. This complicates this picture, but it may be good news as superconductivity could lead to larger asymmetries.  A simple estimate for a type II
superconducting core gives \cite{2002PhRvD..66h4025C,2008MNRAS.383.1551A}
\begin{equation}
\epsilon \approx 10^{-9}
\left(\frac{B}{10^{12} \rm \, G}\right)
\left(\frac{H_{\rm crit}}{10^{15} \rm \, G}\right) \ ,
\end{equation}
where $H_{\rm crit}\approx {10^{15} \rm \, G}$ is the so-called critical field \cite{2002PhRvD..66h4025C}. The main problem is that we do not really know what the internal magnetic field configuration may be. To make matters worse, it seems that most of the models we can build are not actually stable \cite{lander}. 

What do these estimates mean for observation efforts?  Since the sensitivity of a search increases in
inverse proportion to the detector noise level and as the square root
of the observation time,  a
search over two years with an instrument like the Einstein Telescope may be able to 
detect deformations at the $\epsilon \sim 10^{-9}$ level in some of the millisecond pulsars.
Hence, the deformation associated with a typical pulsar magnetic field, for which $\epsilon
\approx 10^{-12}$, is too small to ever be detected.  We need nature to be less conservative than these estimates.

The  challenge is to provide 
reasonable scenarios that lead to the development of sizeable
deformations.  In this sense, accreting systems are promising \cite{1998ApJ...501L..89B} because
of the expected asymmetry of the accretion flow near the star's
surface. Hence, it is not surprising that neutron
stars in low-mass X-ray binaries have attracted considerable attention. 
In fact, the currently
observed spin distribution in these systems seems to suggest the
presence of a mechanism that halts the spin-up due to accretion \cite{patruno}.
Gravitational-wave emission could provide a balancing torque if the accretion leads
to deformations in the crust 
\cite{1998ApJ...501L..89B, 2009MNRAS.395.1972V,2005ApJ...623.1044M}, and it is easy to show that the required
deformation is  small enough to be allowed (at least in principle). Unfortunately,  accreting
systems are messy and we do not understand the detailed
accretion torque very well \cite{2005MNRAS.361.1153A}. While we have interesting information from precision X-ray
timing  we do not yet have a consistent theoretical model for these systems. Despite decades of effort we can not say with certainty that  gravitational waves have a role to play.

\section{The r-mode instability}

A neutron star has a rich oscillation spectrum, intimately linked to the internal composition and state of matter. The star's oscillation modes may be associated with a distinct gravitational-wave signature. We have already touched upon an example of this -- the wild oscillations of a binary merger remnant. In principle, the dependence of the various modes on specific physics aspects may be ``inverted'' to provide us with information that is difficult to obtain in other ways. The basic strategy for such ``gravitational-wave
asteroseismology'' is clear \cite{ak98}, but our models need to be made much more realistic if the method
is to be used in practice. We also need to establish why various oscillation modes
would be excited in the first place and understand what level of excitation one would expect. This problem is challenging because it involves astrophysics that we barely understand even at the back-of-the-envelope level. However, we know that there are  scenarios where oscillation modes may grow large. Neutron stars may exhibit  a number of  instabilities. These include the  instability of the f- and r-modes -- where it is the emission of gravitational waves that drives the instability, the
dynamical bar-mode and low $T/W$ instabilities, there are specific instabilities associated with a relative
flow in a superfluid core etcetera. In recent years our understanding of these instabilities has improved
considerably, but we are still far away from reliable predictions.

As far as instabilities are concerned, the one associated with the 
r-modes \cite{1998ApJ...502..708A,2001IJMPD..10..381A} remains (after nearly 20 years of scrutiny) the most promising. The r-modes belong to a large class of (inertial) modes which are restored by the Coriolis force, and they 
are peculiar in that they are driven unstable by the emission of gravitational radiation already at relatively modest rotation rates (in fact, as soon the star is set into rotation if we ignore the effects of viscosity). In addition to being interesting in its own right, the r-mode problem provides a useful illustration of the intricate interplay between different aspects of neutron star physics required in any ``realistic'' model.

The r-mode instability depends on a balance between gravitational-wave driving (primarily through current multipole radiation) and various dissipation mechanisms. In effect, the instability provides a  probe of the core physics (including aspects of the weak interactions, which determine reaction rates and hence bulk viscosity damping). As an illustration, let us consider a simple model of a neutron star composed
of neutrons, protons and electrons, ignoring issues to do with the crust physics, 
superfluidity, magnetic fields etcetera. If we take the overall density profile to be that of a polytrope then the characteristic
growth timescale for the quadrupole r-mode is \cite{2001IJMPD..10..381A}
\begin{equation}
t_{\rm gw} \approx  50  \left( {1.4 M_\odot \over M } \right) \left( {10\ \mathrm{km} \over R} \right)^4 \left( {1~\mathrm{kHz}\over f_\mathrm{spin} } \right)^6 \ \mbox{s} 
\end{equation}
That is, in a rapidly spinning star the instability grows on a timescale of minutes, much faster than other evolutionary processes.
In the simplest model the unstable mode is damped by  shear and bulk viscosity.
At relatively low core temperatures (below a few times $10^9$~K) 
the main viscous dissipation mechanism  arises from
momentum transport due to particle scattering,  modelled as a
macroscopic shear viscosity. In a normal fluid star neutron-neutron 
scattering 
provides the most important contribution. This leads to a 
typical damping time
\begin{equation}
t_{\rm sv} \approx 7\times10^7 \left( {1.4 M_\odot \over M } \right)^{5/4} \left( { R \over 10\ \mathrm{km}} \right)^{23/4} \left({ T \over 10^9\ \mathrm{K} } \right)^2 \mbox{ s}
\label{sv1}\end{equation}
In other words, at a core temperature of $10^9$~K the damping timescale is 
longer than a year. In a superfluid star the shear viscosity is mainly due to electron-electron scattering, but this does not lead to a very different damping timescale. 
At high temperatures
bulk viscosity is the dominant dissipation
mechanism. Bulk viscosity arises as
the mode oscillation
drives the fluid out of beta equilibrium. The efficiency of this mechanism depends on the extent to which energy is
dissipated from the fluid motion as weak interactions try to re-establish equilibrium. It is essentially a resonant mechanism,  
particularly efficient when the oscillation timescale is similar to the reaction timescale. 
At higher and lower frequencies (or, equivalently, temperatures) the bulk viscosity mechanism is weaker.
The bulk viscosity 
damping timescale is approximately given by
\begin{equation}
t_{\rm bv} \approx 3 \times10^{11} \left( { M \over 1.4 M_\odot} \right) \left({ 10\ \mathrm{km} \over R} \right)\left( {1~\mathrm{kHz} \over f_\mathrm{spin} } \right)^2 \left({ 10^9\ \mathrm{K} \over T} \right)^6
\mbox{ s}
\label{bulkest}\end{equation}
It is easy to see that, in order for this damping to be efficient we need the star to be very hot $\sim10^{10}$~K. This means that the bulk viscosity plays little role for cold mature neutron stars.

\begin{figure}[h]
 \centerline{\includegraphics[width=6cm]{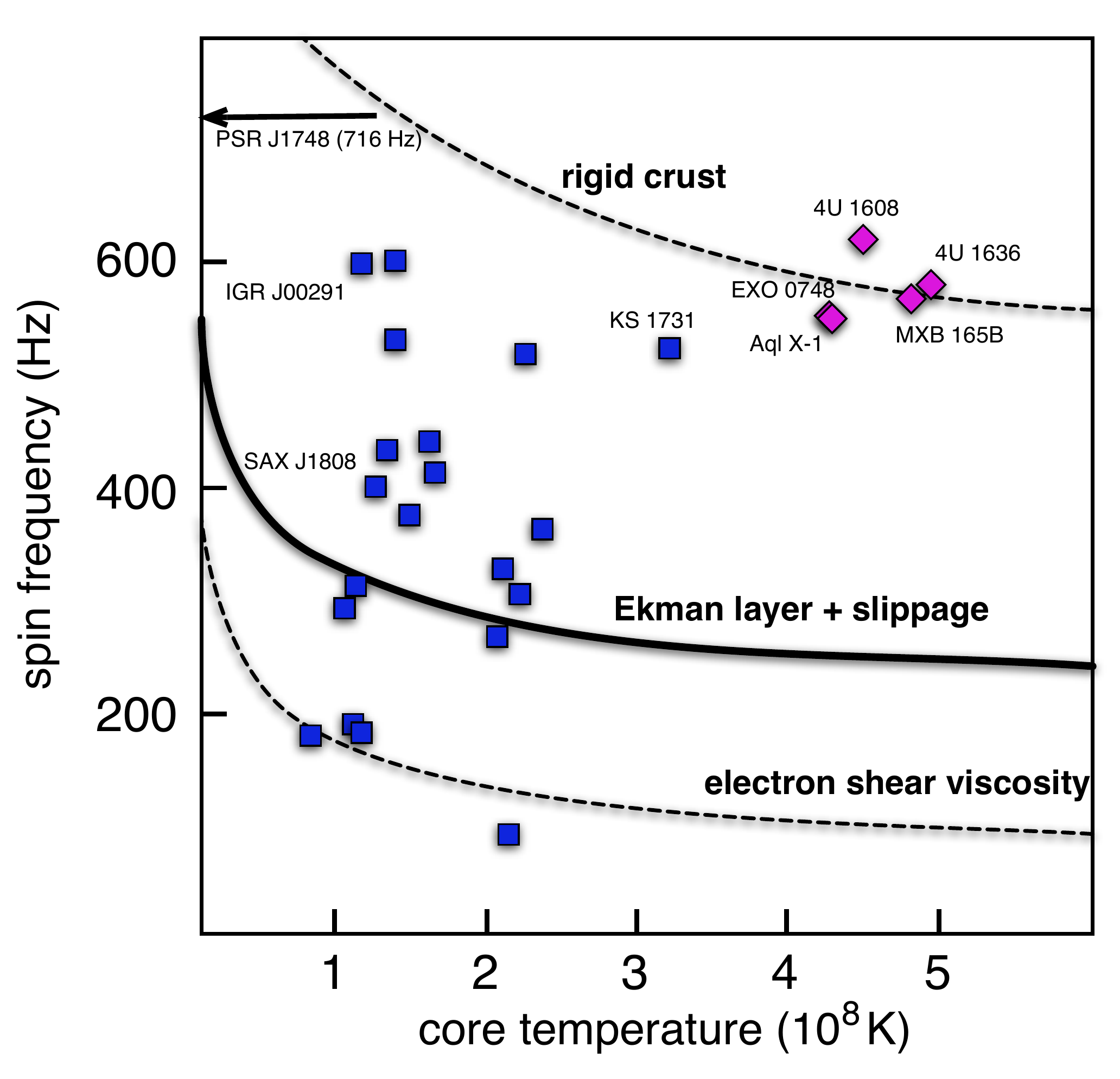}}
  \caption{Comparing different models for the r-mode instability window at low temperature to observed neutron stars in low-mass X-ray binaries. According to the ``best'' current theory, several systems should be unstable but there is no real evidence for this from the observed spin behaviour. Adapted from \cite{hor}.}
\label{rmodes}
\end{figure}

From these estimates we learn that the r-mode
instability will only be active in a certain temperature range.
To have an  instability we need
$t_{\rm gw}$ to be  smaller than both $t_{\rm sv}$ and 
$t_{\rm bv}$. Shear viscosity suppresses the instability at core temperatures below $10^5$~K.
Similarly, bulk viscosity will prevent the mode from growing 
in a star that is hotter than a few times
$10^{10}$~K. In the intermediate temperature range, the growth time due
to gravitational radiation is short enough to overcome the
viscous damping and  the mode is unstable above some critical spin rate.

The basic picture of the r-mode instability has not changed much since the early work on the problem \cite{1998PhRvL..80.4843L,1999ApJ...510..846A}. Different aspects have been considered, sometimes leading to more complicated instability windows, but the basic picture remains the same. As an example of the discussion we may consider the role of the star's crust. If the crust were rigid, then the fluid motion associated with the mode would rub against its base. This would lead to a very efficient damping through a viscous boundary layer (an Ekman layer) \cite{gregus}. However, the crust not solid -- it is more like a jelly -- and at the rotation rates we are interested in the Coriolis force dominates the elasticity. This means that the crust partakes in the r-mode oscillation, and the viscous damping is reduced by a fairly large factor that encodes to what extent the fluid slips relative to the crust \cite{slip}. We can compare the predictions to known accreting neutron stars in low-mass X-ray binaries, for which we know the spin and we also have upper limits on the temperature. Such a comparison is shown in figure~\ref{rmodes}. The message here is clear. If we believe the slippage argument, then many of the observed systems should be r-mode unstable. However, if they were, then the loss of energy through the emitted gravitational waves should spin the stars down. They would not be able to remain inside the instability region. However, there is no (real) evidence for this happening so we are led to conclude that our understanding of r-mode damping is incomplete. Various suggestions have been made to resolve this problem \cite{hor,kg}, but it is probably fair to say that they all involve some level of fine tuning. We are still looking for a convincing explanation. Perhaps it is the case -- as in much of astrophysics -- that the answer involves ``the magnetic field''. We just need to figure out how...

This is  true for other aspects of the r-mode instability, as well. In addition to working out if the modes are likely to be unstable in a given system, we need to know why the unstable modes stop growing -- why the amplitude saturates -- and at what level this happens. We also need to know how a star with an 
 active instability evolves. How does its spin change and at what rate does it heat up/cool down. It is frustrating to admit, after two decades of thinking about these questions, that we do not have good answers. We think we know that the modes saturate as they couple to a sea of short-wavelength inertial modes \cite{arras}, and we expect  the associated spin-evolution to be very complicated \cite{bonda}. This would be quite intuitive (as the mechanism is similar to the early onset of turbulence), but the problem is subtle and the coupling of different modes should (somehow) depend on the detailed physics. The question is if we can take further steps towards realism. Given the complexity of the problem, this would involve an awful lot of work. Again, the  theorists may have to sit back and hope that observers come to the rescue.

\section{Final remarks}

As we sober up from celebrating the first detections of gravitational waves, it is natural to ask what happens next. Obviously, LIGO -- now joined by Virgo -- are taking more data. The expectation is that there will be further black holes signals and the hope is that we will see/hear more than this. Neutron star signals have to be top of the wish list. Given what we (think we) know we should be able to catch them. This would be very exciting and could also be of great importance for the quest to understand physics beyond the laboratory. Neutron star signals would allow us to probe the very extremes of physics. In order to help the detection effort we need to improve our understanding of the theory, but  it may be unrealistic to expect that we will resolve the various involved issues any time soon. Most likely, we need observational data to  constrain  different proposed  mechanisms. As part of this process it is  natural to consider the next generation of detectors (like the Einstein Telescope),  for which key design decisions still have to be made. Perhaps most importantly, we need to keep in mind that the relevant problems are intricate and there is much we do not yet know. We are trying to distinguish  faint whispers compared to the roar of colliding black holes and there is no reason why this should be easy. 

\appendix

\section*{Acknowledgement}

Support from the STFC in the UK is gratefully acknowledged.



\end{document}